\documentclass[aps,prl,a4paper,twocolumn,groupedaddress]{revtex4-1}
\pdfoutput=1
\usepackage[]{latexsym,amsmath,amssymb,graphicx,epsfig}

\newcommand{\be}{\begin{eqnarray}}
\newcommand{\ee}{\end{eqnarray}}
\newcommand{\degree}{\ensuremath{^\circ}}

\begin{document}


\title{Quantum Black Holes from Cosmic Rays}


\author{Xavier Calmet}
\email[]{x.calmet@sussex.ac.uk}
\affiliation{Physics and Astronomy, University of Sussex, Falmer, Brighton, BN1 9QH, UK}

\author{Lauretiu Ioan Caramete}
\email[]{lcaramete@spacescience.ro}
\affiliation{Institute of Space Science, P.O.Box MG-23, Ro 077125 Bucharest-Magurele, Romania}

\author{Octavian Micu}
\email[]{octavian.micu@spacescience.ro}
\affiliation{Institute of Space Science, P.O.Box MG-23, Ro 077125 Bucharest-Magurele, Romania}


\date{\today}

\begin{abstract}
We investigate the possibility for cosmic ray experiments to discover non-thermal small black holes with masses in the TeV range. Such black holes would result due to the impact between ultra high energy cosmic rays or neutrinos with nuclei from the upper atmosphere and decay instantaneously. They could be produced  copiously if the Planck scale is in the few TeV region. As their masses are close to the Planck scale, these holes would typically decay into two particles emitted back-to-back. Depending on the angles between the emitted particles with respect to the center of mass direction of motion, it is possible for the simultaneous showers to be measured by the detectors.  
\end{abstract}


\maketitle

{\em Introduction:\/}
It is now well appreciated that the energy scale at which quantum gravitational effects become important could be anywhere between the traditional Planck scale, i.e. some $10^{19}$ GeV and a few $10^3$ GeV. Brane world models with a large extra-dimensional volume \cite{arkani,RS} or with a large hidden sector of particles in 4 dimensions \cite{Calmet:2008tn}, illustrate that quantum gravity effects can be important in the few TeVs region.

One of the most exciting implications of low scale quantum gravity model is that small black holes could be produced in the collision of particles with center of mass energies above the Planck scale. The classical production of black holes in the collision of two highly boosted objects has been studied both for zero and non-zero impact parameters by Penrose in the seventies, but he never published his results. The state of the art can be found in more recent seminal papers by D'Eath and collaborators \cite{D'Eath:1992hb}  and by Eardley and Giddings \cite{Eardley:2002re}. Remarkably, these authors established the formation of a closed trapped surface in such collisions which is a real mathematical tour de force. This work has been extended to the semi-classical regime by Hsu \cite{Hsu:2002bd}. Small black holes with masses about 5 to 20 times larger than the Planck scale (see \cite{Meade:2007sz} for a recent discussion) are accurately described by semi-classical methods.

Most up to date studies of the production of small black holes at colliders or in cosmic ray collisions have considered semi-classical black holes \cite{Dimopoulos:2001hw,Banks:1999gd,Giddings:2001bu,Feng:2001ib,Anchordoqui:2003ug,Anchordoqui:2001cg,Anchordoqui:2003jr}. Given the argument mentioned above, it is however clear that the number of semi-classical black holes produced at the LHC would be very small even if the Planck mass was at a TeV.
There was a motivation \cite{Calmet:2008dg,Calmet:2011ta,Calmet:2012cn} to consider quantum black hole which are non-thermal objects with masses close to the Planck mass. Because they are non-thermal they are expected to decay only to a few particles, typically two. This implies that quantum black hole signatures are very different from semi-classical objects which are expected to decay into several particles in a final explosion, see e.g. \cite{BHreview} for recent reviews. For the quantum black hole model that we have in mind we do not expect a remnant  and we expect the quantum black holes to decay instantly. In the models considered in \cite{bhEarth1} , there is parameter space for both long lived quantum black holes and for black holes which decay instantaneously like it is the case here. If this signature is observed it will not validate one model or the other but only the existence of TeV range quantum black holes. 

 Besides being produced at colliders, quantum black holes could also be produced in high energetic collisions of cosmic rays with nuclei in the high atmosphere. Cosmic ray experiments might be able to detect such spectacular events. This is the topic of this Letter. While at the LHC the detectors allow for the accurate measurement of the energy balance of the collisions, cosmic ray experiments have another advantage, namely that the center of mass energy for the collisions between ultra high energy cosmic rays or neutrinos with nuclei from the atmosphere can be several orders of magnitude above the maximum energy attainable by the LHC. The disadvantage is that not all of the energy resulting from a collision can be directly measured. All that is measured is the ground imprint of the particle showers and eventually the fluorescence shapes of the showers. The possibility still exists for quantum black holes which decay instantaneously into two particles to be detected by these experiments.  In the lab frame, the secondary showers develop at an angle which can be calculated. If this angle is large enough, the ground detectors of cosmic ray experiments can measure an event for which the ground imprint is formed by two coincidental spatially separated showers. 
 

{\em Black holes production:\/}
 The cross section $\nu$ N $\to$ BH is given by
\begin{eqnarray}
\label{CSection}
\sigma(E_\nu,x_{min},M_D)&=&
\int^1_0 2z dz \int^1_{\frac{( x_{min} M_D)^2}{y(z)^2 s_{max}}} dx F(n)  \\ && \nonumber \pi r_s^2(\sqrt{\hat s},M_{D}) \sum_i f_i(x,Q)
\end{eqnarray} 
where $M_D$ is the $4+n$ dimensional reduced Planck mass, $z=b/b_{max}$, $x_{min}=M_{BH,min}/M_D$,  $n$ is the number of extra-dimensions, $F(n)$ and $y(z)$ are the factors introduced by Eardley and Giddings \cite{Eardley:2002re} and by Yoshino and Nambu \cite{Yoshino:2002tx}. The $4 + n$ dimensional  Schwarzschild radius is given by 
\begin{eqnarray}
r_s(us,n,M_D)=k(n)M_D^{-1}[\sqrt{us}/M_D]^{1/(1+n)}
\end{eqnarray}
where
\begin{eqnarray}
k(n) =  \left [2^n \sqrt{\pi}^{n-3} \frac{\Gamma((3+n)/2)}{2+n} \right ]^{1/(1+n)}.
\end{eqnarray}
Furthermore, note that $\hat{s}= 2 x m_N E_{\nu}$ where $m_N$ is the  nuclei mass and $E_{\nu}$ is the neutrino energy. The functions $f_i(x,Q)$ are the parton distribution functions.  We note that black hole production due to cosmic neutrinos might be suppressed \cite{Stojkovic:2005fx}, but this is a model dependent question.
For proton nuclei collision we have
\begin{eqnarray}
\sigma^{pN}(s,x_{min},n,M_D) &=& \int_0^1 2z dz \int_{\frac{(x_{min} M_D)^2}{y(z)^2 s}}^1 du \\ \nonumber && \times
 \int_u^1 \frac{dv}{v}  F(n) \pi r_s^2(us,n,M_D)
\\
\nonumber && \times
 \sum_{i,j} f_i(v,Q) f^N_j(u/v,Q)
\end{eqnarray}

The number of black holes expected to be seen by a cosmic rays experiment is given by
 \begin{eqnarray}
N&=&  \int dE N_A  \frac{d\Phi}{dE} \sigma(E) A(E) T
\label{N}
\end{eqnarray} 
where  $\sigma(E)$ is the relevant production cross section given above, $\frac{d\Phi}{dE_\nu}$ is the flux of cosmic ray particles, $A(E)$ is the acceptance of the experiment in cm$^2$ sr yr, $N_A$ is the Avogadro number and $T$ is the running time of the detectors.

The possible sites for accelerating particles at very high energy include Active Galactic Nuclei (AGN) and Gamma Ray Bursts (GRBs). Assuming that one of them is the dominant class of source for the ultra high energy cosmic rays (UHECR), predictions can be made about the all sky flux and energy spectrum \cite{Caramete:2011vd}. This will provide a flux of particle dominated in principal by protons.

Another stream of energetic particles is represented by high energy neutrinos which can be estimated by considering two sites of productions: at the source and between the source and the detection place, usually Earth. The production sites of the extragalactic UHECR, which include (AGN) and (GRBs), are currently associated with the ones for neutrinos which are produced through pion decay in proton-proton or proton-photon interactions within the source \cite{Olinto:2000sa}.

The flux of cosmogenic neutrinos depends on the composition of the cosmic rays at high energies, which can be either protons, neutrons, heavy nuclei or a combination of these \cite{Allard:2006mv,Hooper:2004jc}.
The interaction of nuclei with the background does not directly lead to any neutrino flux, neutrinos are produced in the decay of neutrons as products of the dissociation of the nuclei. On the other hand protons produce high energy neutrinos through pion production. The neutrino flux produced by the nuclei has a characteristic maximum at much lower energy than the one due to protons.

In the following the focus will be on this two classes of particles, protons and neutrinos as main particles responsible for creating quantum black holes in the atmosphere of the Earth.

{\em Black holes signature:\/}
The first informations needed to estimate how the showers resulting from the decay of the black holes develop and the possible signatures are the mass of the black holes - $M_{BH}$ and the Lorentz factor $\gamma_{BH}$. Note that the mass of the quantum black hole is dialed by the energy of the incoming particles and is a continuous quantity. Extensive work has been done towards calculating the black hole mass for the case of two particles moving towards one another with similar energies, like it is the case at the LHC. The Lorentz factors of the resulting black holes in these instances are several orders of magnitude smaller in comparison to those studied in the present article. 

To describe the process accurately, one needs to take into account the amount of energy which is radiated via gravitational radiation, the dependence of the horizon formation and black hole mass on the value of the impact parameter $b$ which is defined as the perpendicular distance between the paths of the two colliding particles. For the purpose of this letter knowing the orders of magnitude for $M_{BH}$ and $\gamma_{BH}$ is sufficient. If the impact parameter is small enough for a black hole to form, the mass and Lorentz factor of the resulting black hole vary by less than an order of magnitude when considering all the above mentioned effects. Therefore, the events of black hole creation are described using a simple relativistic calculation starting from the assumption that the entire energy of the two  particles, i.e. the partons of the protons and/or neutrinos goes into the black hole creation. 


A black hole produced as a result of the collision between an UHECR of mass $m_1$ moving relativistically with $\gamma_1$, and a particle $m_2$ at rest has a mass
\be
M=\sqrt{m_1^2+m_2^2+2\gamma_1 m_1 m_2}~,
\label{mbh}
\ee
and is moving relativistically with 
\be
\gamma=\frac{\gamma_1 m_1+m_2}{M_{BH}}~.
\label{pbh}
\ee
The results represent the mass and Lorentz factor for the case of a "sticky" collision between two particles. 
The equations hold regardless of the mass of the resulting object, but a black hole can form only when $M>M_{Pl}$. Assuming this condition to be satisfied, for the rest of the article $M_{BH}$ and $\gamma_{BH}$ will be used when referring to black hole mass and Lorentz factor. 
A similar exercise can be performed to calculate the masses and Lorentz factors for black holes which are formed by the collisions of high energy neutrinos with particles in the atmosphere.  Working under the approximation of massless neutrinos, the black hole mass and Lorentz factor are found to depend on the neutrino energy and the mass of the particle that they collide with. However, when the incident neutrino or UHECR have the same energy, the resulting black holes have similar masses and Lorentz factors.

Quantum black holes are non-thermal objects (their mass is close to the Planck mass). Therefore they are expected to decay to only a couple of particles.
The two particles with masses $m_a$ and $m_b$, produced by the instantaneous decay of a quantum black hole, are emitted back-to-back in the center of mass and with no preferred direction since the differential cross section is angle independent. The only restriction in this case is for the sum of the two masses  $m_a$ and $m_b$ to be smaller than $M_{BH}$.
In the center of mass, the momenta of the two particles are opposite vectors with magnitudes equal to
\be
\!\!\!p\!=\!\frac{\left[\left( M_{BH}^2-(m_a+m_b)^2\right)\!\!\left( M_{BH}^2-(m_a-m_b)^2\right)\right]^{\frac{1}{2}}}{2M_{BH}}.
\ee

One can calculate the energies and momenta of the two particles in the laboratory reference frame (Earth reference frame) by using the Lorentz transformations for the momentum 4-vector
 \be
        \left(\begin{array}{c} E_i' \\ p_{i \parallel}' \end{array}\right) &=&
        \begin{pmatrix} \gamma_{BH} & -\beta_{BH} \gamma_{BH}  \\ -\beta_{BH} \gamma_{BH}  & \gamma_{BH} 
         \end{pmatrix}  \label{LT}
        \left(\begin{array}{c}  E_i \\ p_{i \parallel} \end{array}\right)\\
      p_{i \perp}' &=&p_{i \perp} \nonumber
    \ee
where $i=a, b$; $E_i$ and $p_i$ are the center of mass energy and momentum for the $i$-th particle, while the primed quantities are the corresponding ones measured in the lab reference frame. The 3-momentum ${\bf p}$ of each particle was decomposed into two components: $p_{i\parallel}$ - the momentum component parallel to the direction of motion of the center of mass and $p_{i\perp}$ - the momentum component perpendicular to this direction.

Assuming that the angles of motion for the two resulting particles in the center of mass reference frame with respect to the direction of motion of the center of mass are $\phi_a$ and $\phi_b$, where $\phi_a + \phi_b=\pi$ (the particles are moving back-to-back), one can use the Lorentz transformations from Eq. \ref{LT} to calculate the values of these two angles in the lab reference frame using
\be
\tan {\theta_i}=\frac{\sin{\phi_i}}{\gamma_{BH} \beta_{BH}\frac{E_i}{p_i}+\gamma_{BH} \cos{\phi_i}}~,
\label{tangent}
\ee 
These are the angles between the secondary showers and the direction of motion of the center of mass.

Such a distinctive black hole decay signature can be observed if the angular separation between the secondary showers is large enough for the experiment to be able to resolve the event in two distinctive coincident showers. Numerical simulations show that there is parameter space for large angular separation between the two secondary showers. 

In the following we shall consider quantum black holes created as a result of the collisions between UHECRs or neutrinos with energies larger than $10^{6}$ TeV with particles in the atmosphere. This is the range of energies which are visible to the Pierre Auger Observatory. The Planck scale is assumed to be around $10$ TeV and black holes can form with masses anywhere above this scale.
As stated before, quantum black holes (having a mass of one to five Planck masses) are non-thermal and decay back-to-back into two standard model particles.  Depending on the particles which collide to form the black holes, they can carry various standard model charges and their decay channels will be different. In this letter we do not go to such depth and analyze particular decay channels, but we are only interested in the dependence of the separation between the two resulting showers on the masses of the particles that the black holes decay into. 

\begin{figure}
\includegraphics[scale=0.7]{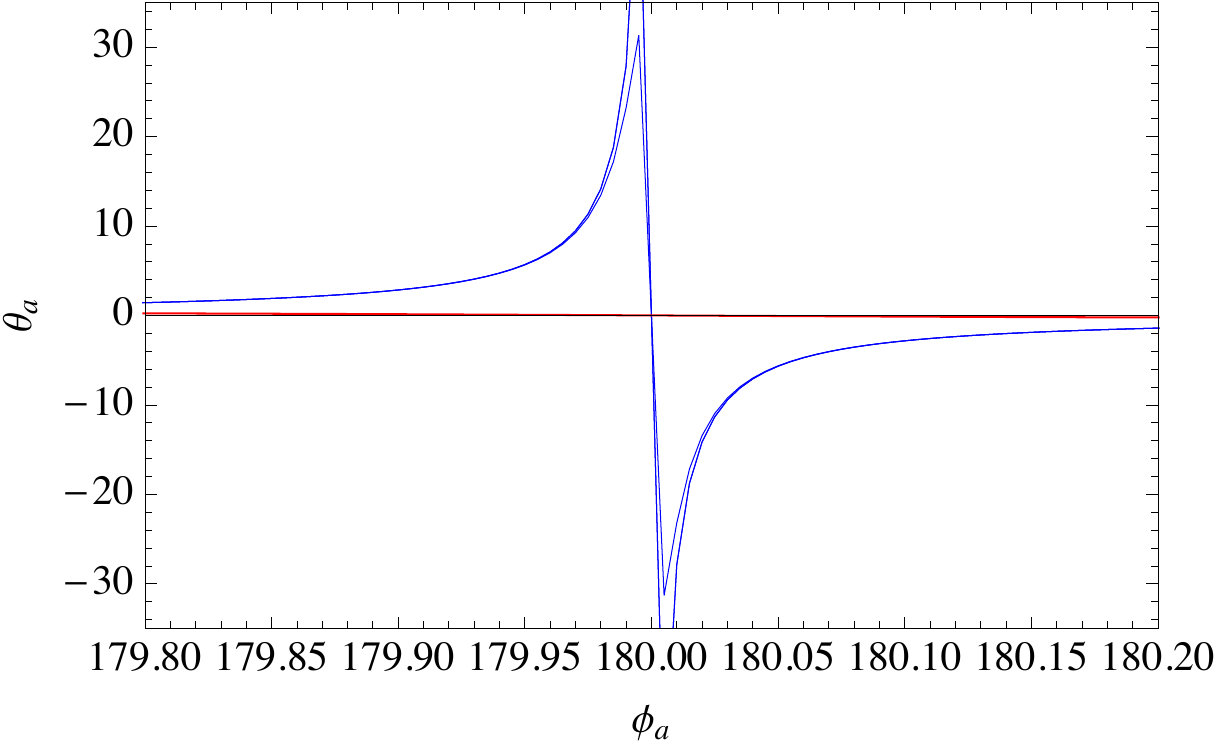}
\caption{Angle $\theta_a$ as  function of the angle $\phi_a$ for $m_a$ corresponding to $m_d= 4.8$ MeV, $m_{\mu}= 105.7$ MeV, $m_{\tau}= 1.777$ GeV, $m_t= 171.2$ GeV (decreasing from a larger possible angle for the lowest value of $m_a$, to a lower possible one for the largest value of $m_a$. The particles for which the angle $\theta_a$ goes through a large maximum value are represented in blue (down quark, muon, tau), while the particles for which $\theta_a$ remains close to $0 \degree$ are represented in red (top quark).} 
\label{plot1}
\end{figure}

For the numerical simulations we consider several standard model particles which can result from the back-to-back decay of quantum black holes. Our choice is such that their masses cover a range of several orders of magnitude: down quark ($m_d= 4.8$ MeV), muon ($m_{\mu}= 105.7$ MeV), tau ($m_{\tau}= 1.777$ GeV), top ($m_t= 171.2$ GeV). Depending on the standard model charges which the black holes carry one can infer the second particle ($m_b$) which results from the back-to-back decay. Fig. \ref{plot1} is a plot of the angle $\theta_a$ as a function of $\phi_a$ for each of the four particles taken into consideration. Only the range of values for which the angle of separation between the direction of motion of the center of mass and the direction of the center of the secondary shower in the laboratory reference frame is large is plotted. The total angle between the centers of the two secondary showers resulted by the decays of the two particles is the sum of the angles $\theta_a$ and $\theta_b$, angles which can be calculated using Eq. \ref{tangent}. However when one of the angles increases as shown in the figure, the other angle decreases to almost zero, and the sum can be very well approximated by the larger of the two. 
As far as the dependency of the maximum possible value of the angle of separation $\theta_a$ on the mass of the particle $m_a$ is concerned, an inverse proportional dependence is observed. The smaller the mass $m_a$, the larger the angle of separation between the two showers can get. It can be seen in Fig. \ref{plot1} that for the heaviest particle considered, the top quark (red curve), the angle $\theta_a$ does not go through a peak value but is approximately zero throughout the entire interval. This is directly related to the fact that the Lorentz factors in the center of mass for the heavier particles are smaller and this in turn has an effect on the corresponding quantities when measured in the laboratory reference frame. Even if the masses of the the two initial particles can be neglected when numerically simulating the development of the showers, the masses are important when calculating the direction/separation of the showers. The discovery of this black hole decay signature is based on the existence of a spatial separation between the centers of the two secondary showers. This is why one has to take the masses of the particles which result from the back-to-back decay into consideration in this analysis. 

Assuming that the showers are created in the atmosphere about  50 km above the Earth, a separation angle (in the laboratory reference frame) larger than $1\degree$ leads to a separation between the air shower axes at Earth level of at least 1 km, leading to a positive detection of this signature by present cosmic ray observatories.
The plot in Fig. \ref{plot1} shows that only for a range of values of the angle $\phi_a$ covering about $0.4 \degree$ as measured in the center of mass reference frame ($179.8\degree<\phi_a<180.2\degree$) can the angle of separation between the two secondary showers (measured in the experiment reference frame) be large enough to be observed experimentally.

\begin{figure}
\includegraphics[viewport=2cm 4cm 20cm 29cm,clip,scale=0.48]{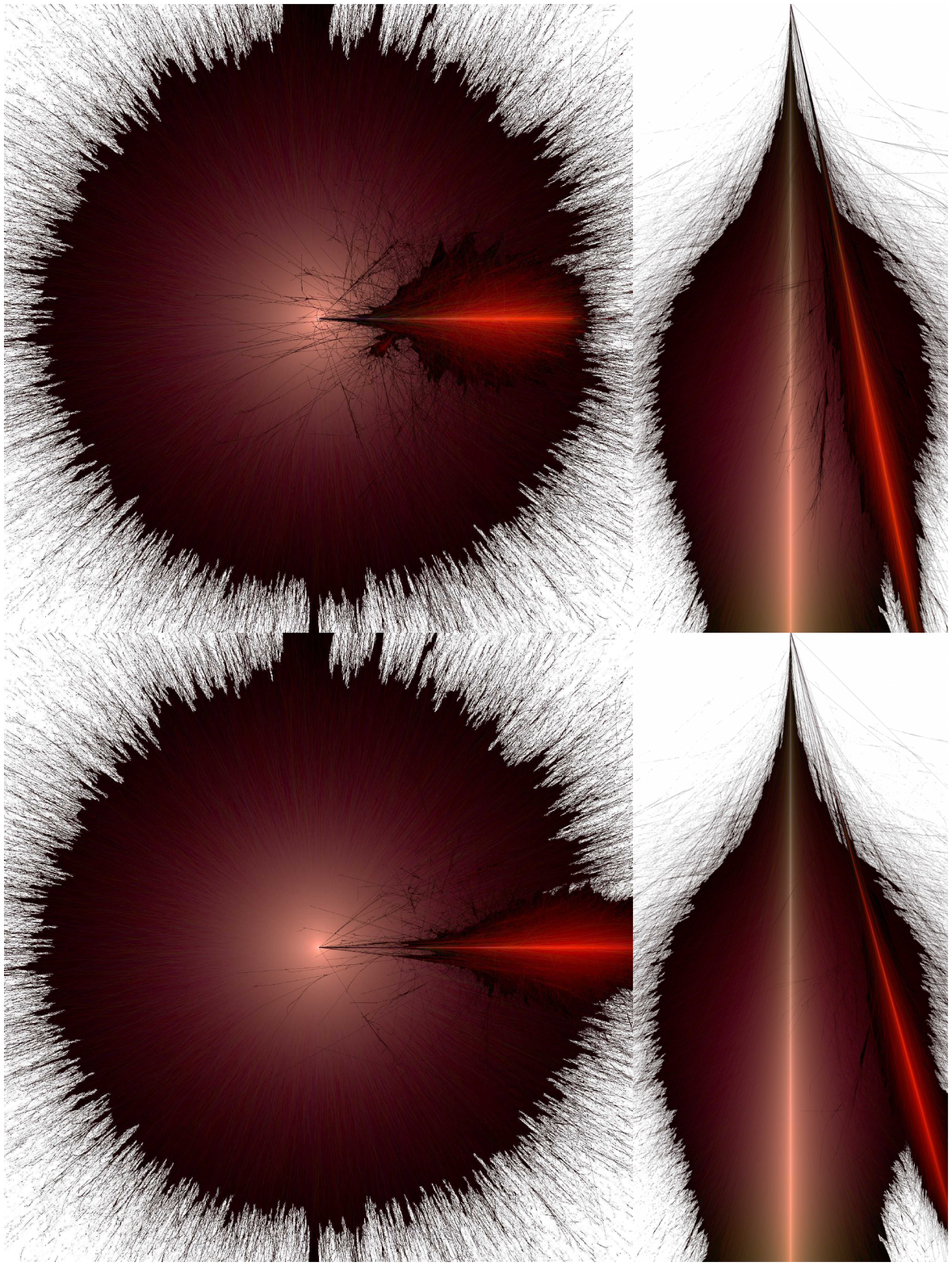}
\caption{CORSIKA simulations of two showers with energies of $10^{13}$ eV and $9\times 10^{16}$ eV developing from 30 Km above the ground. The two showers develop from the particles resulting from the back-to-back decay of a black hole produced by an UHECR having an energy of $10^{17}$eV. The large shower develops vertically. The angle between the showers are $15 \degree$ (top) and $20 \degree$ (bottom). Images on the left represent simulation for the showers as viewed from the top and on the right the sowers as viewed from the side. } 
\label{plot2}
\end{figure}

The signatures of back-to-back black hole decays can also be detected when the two showers overlap partly resulting in oval shaped imprints being seen by the ground detectors as it is shown in the numerical simulations presented in Fig. \ref{plot2}. The simulations were obtained using CORSIKA-6990 \cite{corsika, corsika1}, which is a program used for extensive air showers simulations initiated by high energy cosmic ray particles. One can distinguish between black hole decay events (which leave oval ground imprints) and standard ones (which happen due to the showers coming at an angle with respect to the ground) if there are inconsistencies between the reconstructed directions of the showers using the ground detectors and the showers as seen by the fluorescence detectors. 

Angles close to $0\degree$ and $180\degree$ in the center of mass frame (a necessary condition for the two showers to be well enough separated spatially in the Earth reference frame) also imply large but opposite components of $p_{i\parallel}$ for the two particles. This results in the two showers having very different energies in the Earth reference frame and this has a direct consequence on the size of their ground imprint as seen in Fig. \ref{plot2}. Of course, calculating their initial energies is a standard procedure for cosmic rays experiments. The reason we emphasize this finding is that when the two showers overlap partially, this detail can also contribute to determining how the shower was generated.

\begin{figure}
\includegraphics[scale=0.27]{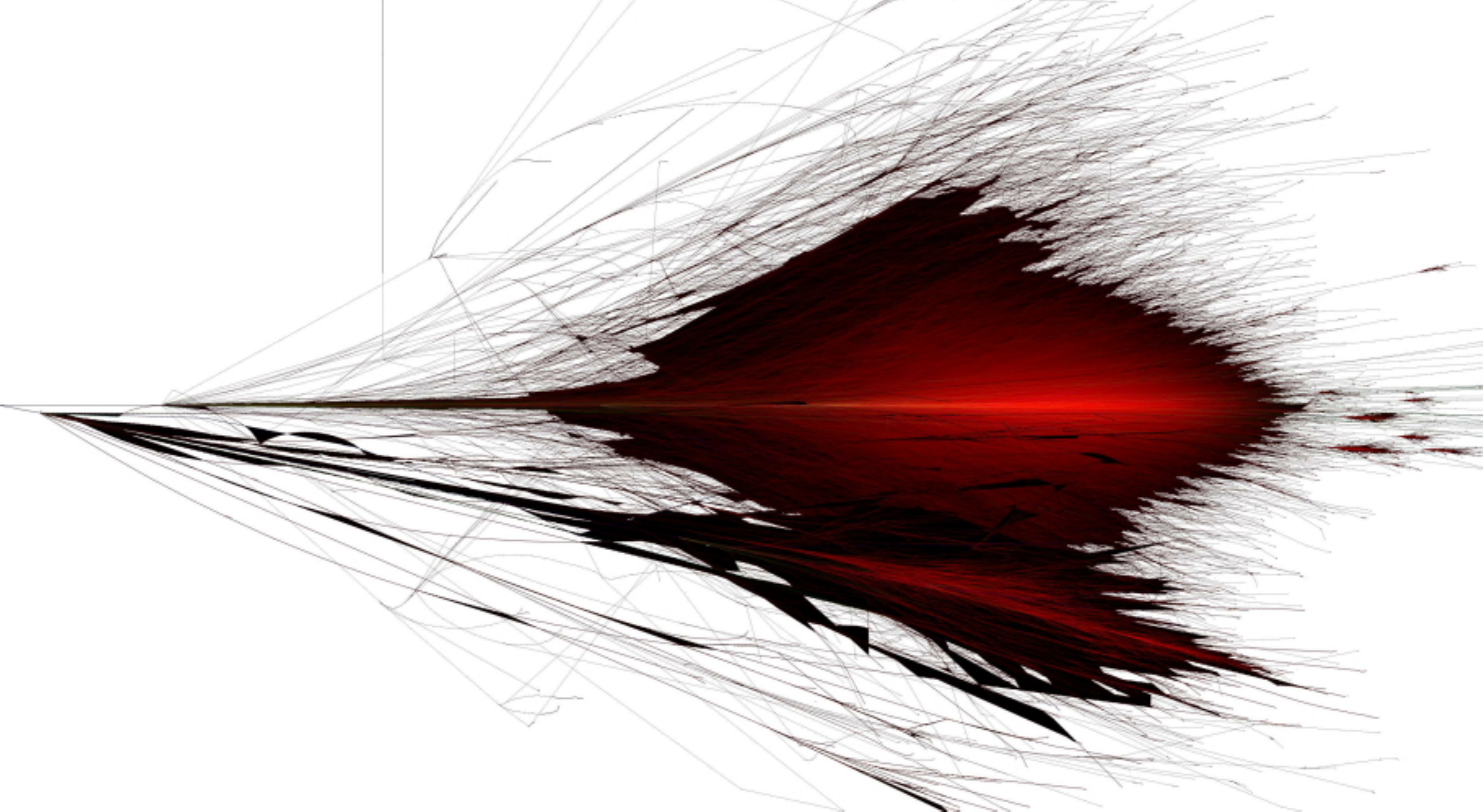}
\caption{CORSIKA simulations of two showers with energies of $10^{13}$ eV and $9\times 10^{16}$ eV developing horizontally in the atmosphere (parallel to the ground). The two showers develop at an angle of $10 \degree$ from the particles resulting from the back-to-back decay of a black hole produced by an UHECR having an energy of $10^{17}$eV.} 
\label{plot3}
\end{figure}

This signature is not only interesting for Earth based cosmic ray experiments but it will actually be easier to be searched for by future space based cosmic ray observatories such as JEM-EUSO (Extreme Universe Space Observatory which will be installed on board of the Japanese Experiment Module on the International Space Station) \cite{Casolino:2011zz}. This fact is obvious when comparing Fig. \ref{plot3} with Fig. \ref{plot2}. From the plots presented in the two figures one can see that the spatial separation of the two showers is not very obvious throughout the entire length of such an event. The more energetic shower can enclose the less energetic one for some parts of the shower development. If the event intersects the plane of the ground in the region where the two showers are not separated, the ground detectors are likely to miss the event. A space observatory would be able to capture the entire development of these events and record something similar to what the simulation in Fig. \ref{plot3} shows. If an event such as the simulation in Fig. \ref{plot3} is found, it would be obvious that it is composed of two distinctive showers which originate in the same point. This capability, along with the much larger acceptance for a space observatory (JEM-EUSO is estimated to have a twenty times larger acceptance than the Pierre Auger Observatory), make a space cosmic ray observatory an ideal candidate for finding this signature.  

Another possibility is for the two showers which develop from the two particles that resulted in the back-to-back decay to arrive to the ground with a measurable time difference between them. The details of this possibility will be investigated in a subsequent article.

One may worry that our signature would be difficult to differentiate from QCD background which can give rise to two high energetic jets as well via processes of the type $q + \bar q \to \mbox{dijets}$ and $q + g \to \mbox{dijets}$. However, the number of QCD events of this type is strongly suppressed by a factor $\alpha_S^2$, where $\alpha_S$ is the QCD coupling constant squared divided by $4\pi$, compared to the gravitationally induced ones. At the scale $M_P\sim$ TeV, $\alpha_S^2 \sim 8 \times 10^{-3}$, we thus expect about 100 events of this type above the QCD background if the Planck mass is at a TeV. Note that for a larger Planck mass, the QCD background is even smaller due to asymptotic freedom.

One might say, in cases where the energy of the cosmic ray is spread roughly equally over the different partons, that the two shower events we described above might be accompanied by showers originating from QCD type reactions among the remaining partons of the colliding cosmic ray and nuclei in the upper atmosphere. This type of events is part of the background discussed above which we anticipate to be suppressed. However, to verify this thoroughly, in a future paper we plan to take the effect of spectator partons into consideration. For this we will use one or several of the available Monte Carlo codes which are capable of doing this part: CHARYBDIS \cite{Harris:2003db}, BlackMax \cite{Dai:2007ki} or QBH \cite{Gingrich:2009hj}.

The phenomenologically interesting cases are for $n=0,~1,~4,~5,~6,~7$ extra dimensions. Note that the case $n=0$ corresponds to the model described in Ref. \cite{Calmet:2008tn} where it is shown that in 3+1 dimensions the Planck scale can be lowered to the TeV region if there is a large hidden sector of particles that interacts only gravitationally with the standard model fields. This model reproduces a lot of the features of extra-dimensional models but obviously does not have Kaluza-Klein excitations of the graviton and is thus far less constrained by current experiments. The $n=1$ case corresponds to the Randall-Sundrum model since for the ADD scenario the case with a single extra dimension is already excluded experimentally. Using the experiment acceptance \cite{Abreu:2011pj} and a fit for the cosmic ray flux \cite{Abraham:2008ru} in Eq. \ref{N}, one can find the total number of quantum black holes (with the mass between one and five Planck masses) created per year by UHECR collisions in the atmosphere to be equal to 11, 81, 460, 609, 765 respectively 925 (for $n=0,~1,~4,~5,~6,~7$). For these numerical estimations a value of  $10$ TeV was assumed for the Planck scale. 
There is no preferential direction in the center of mass along which the decay occurs, and using a simple Monte Carlo simulation one estimates the probability of a back-to-back decay to happen in the $179.8\degree$ to $180.2\degree$ range to be about $0.11~\%$ our of the total number of black hole decay events. 

{\em Conclusions:\/}
We have analyzed the possibility to test the Plank mass in the 10 TeV range and above this value by discovering back-to-back decays of Planck scale quantum black holes in the cosmic ray data. The particular signature for this type of event that we propose to be searched for consists in two simultaneous spatially separated showers pointing to the same origin. It was shown that even if very small, there is available parameter space for this signature to be discovered. The number of expected events varies with the dimensionality of space-time, and when observed, the number of events will be able to point to the correct phenomenological model. As it can be seen from the figure, depending on the values of the masses of the two ``daughter'' particles, the angle of separation can be as high as several tens of degrees. The probability for such events is very small, but the flux of cosmic rays with energies above $10^{6}$ TeV is large enough to make this possible. The article shows that future space based cosmic ray observatories will be even more suitable for these searches. Their increased acceptances will result in a more than ten times larger likelihood of discovery for this type of events when compared to current ground based observatories.

The value of 10 TeV was chosen as an example, but this black hole decay signature can be used to search for any Planck mass value in the range of black hole masses which can be obtained from UHECR collisions with particles in the upper atmosphere. We conclude by emphasizing the importance of this particular black hole decay signature: it allows the cosmic rays experiments to join the LHC efforts to search for TeV scale micro black holes. It even provides them with two advantages: a very distinctive signature and the possibility to search for the Planck scale in a wider range of values than at today's particle accelerators. 
\vspace{1mm}
\paragraph*{Acknowledgements:} This collaboration was made possible by the COST - Action MP0905. OM and LIC were supported by research grants: UEFISCDI project PN-II-RU-TE-2011-3-0184 and LAPLAS 3.

\end{document}